\documentclass[sigconf,screen,authorversion,nonacm]{acmart}

\AtBeginDocument{%
  }

\copyrightyear{2024}
\acmYear{2024}
\setcopyright{rightsretained}
\acmConference[ICSE-SEET '24]{46th International Conference on Software Engineering: : Software Engineering Education and Training}{April 14--20, 2024}{Lisbon, Portugal}
\acmBooktitle{46th International Conference on Software Engineering: : Software Engineering Education and Training (ICSE-SEET '24), April 14--20, 2024, Lisbon, Portugal}
\acmDOI{10.1145/3639474.3640061}
\acmISBN{979-8-4007-0498-7/24/04}

\usepackage{fancyhdr}
\AtBeginDocument{%
	\addtolength{\footskip}{2.0\baselineskip}%
	\fancyfoot[L]{\textit{\textbf{Preprint --- do not distribute.}}}%
}

\usepackage{listings}
\usepackage{graphicx}
\usepackage{verbatim}
\usepackage{balance}
\usepackage[most]{tcolorbox}

\newtcolorbox[auto counter]{recommendation}[1]{title={\bfseries Main Findings for RQ~\thetcbcounter},drop shadow={black!10!white},
  coltitle=white, colframe=white!25!black, boxsep=2pt,left=4pt,right=4pt,top=3pt,bottom=3pt,  sharpish corners}
  
\usepackage{csquotes}
\MakeOuterQuote{"}

\begin{document}

\title{AI-Tutoring in Software Engineering Education}
\subtitle{Experiences with Large Language Models in Programming Assessments}

\author{Eduard Frankford}
\email{eduard.frankford@uibk.ac.at}
\orcid{0009-0005-5959-4936}
\affiliation{%
  \institution{University of Innsbruck}
  \city{Innsbruck}
  \country{Austria}
}

\author{Clemens Sauerwein}
\email{clemens.sauerwein@uibk.ac.at}
\orcid{0009-0009-9464-5080}
\affiliation{%
  \institution{University of Innsbruck}
  \city{Innsbruck}
  \country{Austria}
}

\author{Patrick Bassner}
\email{patrick.bassner@tum.de}
\orcid{0009-0006-0434-6182}
\affiliation{%
  \institution{Technical University of Munich}
  \city{Munich}
  \country{Germany}
}

\author{Stephan Krusche}
\email{krusche@tum.de}
\orcid{0000-0002-4552-644X}
\affiliation{%
  \institution{Technical University of Munich}
  \city{Munich}
  \country{Germany}
}

\author{Ruth Breu}
\email{ruth.breu@uibk.ac.at}
\orcid{0000-0001-7093-4341}
\affiliation{%
  \institution{University of Innsbruck}
  \city{Innsbruck}
  \country{Austria}
}

\begin{abstract}

With the rapid advancement of artificial intelligence (AI) in various domains, the education sector is set for transformation. The potential of AI-driven tools in enhancing the learning experience, especially in programming, is immense. However, the scientific evaluation of Large Language Models (LLMs) used in Automated Programming Assessment Systems (APASs) as an AI-Tutor remains largely unexplored. Therefore, there is a need to understand how students interact with such AI-Tutors and to analyze their experiences. 

In this paper, we conducted an exploratory case study by integrating the GPT-3.5-Turbo model as an AI-Tutor within the APAS Artemis. Through a combination of empirical data collection and an exploratory survey, we identified different user types based on their interaction patterns with the AI-Tutor. Additionally, the findings highlight advantages, such as timely feedback and scalability. However, challenges like generic responses and students' concerns about a learning progress inhibition when using the AI-Tutor were also evident. This research adds to the discourse on AI's role in education.

\end{abstract}


\maketitle

\section{Introduction}
\label{Introduction}

The recent rise of artificial intelligence (AI) has resulted in transformative changes across various sectors. In healthcare, AI has enabled advanced diagnostics and personalized treatments \cite{ho2020enabling}. In finance, algorithmic trading and fraud detection have been revolutionized \cite{ghimire2020accelerating} and the automotive industry is on the brink of a new era with the development of autonomous vehicles  \cite{zhang2021study}.
We have seen first applications of AI in the educational sector through Intelligent Tutoring Systems (ITS) \cite{Crow_2018}.
ITS offer personalized learning experiences, yet their reliance on limited training data confines their applicability to specific scenarios \cite{Crow_2018}. This limitation not only escalates development costs but also restricts the scope and depth of feedback, thus hindering their broader adoption in diverse educational contexts. 

Recently, with the introduction of ChatGPT, we have entered the age of accessible generative AI (GenAI) and large language models (LLMs). LLMs are trained on vast amounts of diverse data and can therefore generate nuanced, comprehensive, and context-aware feedback \cite{lo2023impact}. Beyond just unit test feedback, LLMs like OpenAI's GPT-3.5-Turbo or GPT-4 have the potential to recognize a broader spectrum of student mistakes and offer tailored guidance. Such capabilities can bridge the gap in the shortcomings of traditional ITSs and expand the horizon for feedback mechanisms within programming education.
While the integration of LLMs into various tools and sectors is well-documented, its specific application in programming education, especially, in the form of an AI-Tutor within APASs, remains mostly unexplored.

To address this gap, we seek to address the following research questions:

\begin{enumerate}
    \item \textbf{RQ1:} What is the nature of student interaction with Automated Programming Assessment Systems when facilitated by an AI-Tutor?
    \item \textbf{RQ2:} How do students experience AI driven feedback in Automated Programming Assessment Systems?
    \item \textbf{RQ3:} What are the lessons learned after implementing and operating an AI-Tutor within an Automated Programming Assessment System?
\end{enumerate}

As a first step, the primary objective is to explore the effectiveness and implications of integrating an AI-Tutor based on GenAI, specifically OpenAI's GPT-3.5-Turbo model, into an APAS \cite{keuning2016towards}. 
This approach combines empirical data collection with an exploratory survey. As part of the empirical investigation we closely monitored the AI-tutor's usage, student interactions, code submissions, and feedback timings. Additionally, we analyzed code changes between submissions to understand student engagement patterns with the AI-Tutor.
The preliminary findings suggest that the AI-Tutor offers unique benefits, but there is still a long way to fully optimize the student learning experience.

The remainder of this paper is structured as follows: Section 2 provides an overview of related work. Section 3 elaborates on the research techniques. Section 4 presents the main findings of this study, which are further discussed in Section 5. Section 6 outlines potential constraints of this study, and we conclude in Section 7, summarizing the main insights and reflecting on the broader implications of this research.

\section{Related Work}

ITSs have long been a subject of interest in the realm of programming education \cite{Anderson_1986}. These systems are generally designed to deliver instructional content in a way that is tailored to individual learners, adapting to a student's needs \cite{Crow_2018}. There have been experiments proving that these systems show similar effects like human tutoring \cite{kulik2016effectiveness}. As a result, many ITSs have been created for programming education \cite{Anderson_1986, brusilovsky1992intelligent, holland2009j}. Adaptive or intelligent feedback is a common feature, but this feedback is mainly generated by extensive unit testing \cite{Crow_2018}. 

Beside unit testing, the application of machine learning to emulate human feedback is no recent advancement, as the first chat-bot has been introduced over 50 years ago \cite{bradevsko2012survey}. Since then, these chat-bots have become more and more intelligent \cite{kuhail2023interacting, winkler2018unleashing}. However, they are normally trained on questions the creators expect users to ask, but this is changing with the introduction of ChatGPT \cite{daun2023chatgpt}.

\citeauthor{rudolph2023chatgpt} did one of the first extensive literature reviews on ChatGPT and focused on its relevance for higher education, especially on student assessment, student learning and teaching \cite{rudolph2023chatgpt}. They found that with ChatGPT it is now possible to simulate the assistance provided by a tutor, such as providing personalised assistance in solving problems. 
Furthermore, \citeauthor{ray2023chatgpt} focused on the applications of ChatGPT across various domains and found among other things that it has potential in personalizing learning, by analyzing data on students' learning preferences, strengths, and weaknesses \cite{ray2023chatgpt}. \citeauthor{Kasneci2023chatgpt} discuss the opportunities and challenges when using generative AI tools like ChatGPT in education \cite{Kasneci2023chatgpt}. They point out the opportunity to provide personalized feedback to students.

Literature also already documents the effective use of ChatGPT in improving source code. For example, \citeauthor{surameery2023use} explored the use of ChatGPT to solve programming bugs \cite{surameery2023use}. To be precise, they examined how they can leverage the model to provide debugging assistance, bug prediction and bug explanation to help solve programming problems. They conclude that ChatGPT can play an important role in solving programming bugs, but it is not a perfect solution and should be seen as an additional debugging tool.
\citeauthor{sobania2023analysis} analyzed the automatic bug fixing performance of ChatGPT using the bug fixing benchmark set, QuixBugs \cite{sobania2023analysis}. They found that ChatGPT's bug fixing performance is notably better than other state of the art approaches being able to solve 31 out of 40 bugs.
Other researchers, like \citeauthor{ouh2023chatgpt} and \citeauthor{tian2023chatgpt}, conducted empirical analyses of ChatGPT's potential as a programming assistant focusing on code generation, program repair, and code summarization \cite{ouh2023chatgpt, tian2023chatgpt}. \citeauthor{tian2023chatgpt} found that ChatGPT can hint surprisingly well to the original intention behind what a correct version of a program should look like \cite{tian2023chatgpt}.

\citeauthor{pardos2023learning} concentrated on comparing the efficacy of hints authored by human tutors and hints generated by ChatGPT for elementary and intermediate Algebra \cite{pardos2023learning}. They found that 79\% of hints produced by ChatGPT passed a manual quality test. Additionally, \citeauthor{lo2023impact} conducted research to decide on how ChatGPT performs in different subject domains \cite{lo2023impact} and found that ChatGPT overall performance regarding programming was outstanding to satisfactory \cite{stutz2023ch}. However, regarding "Software Testing", it was able to answer 55.6\% of the questions partially correctly \cite{jalil2023chatgpt}.

Industry has also recognized the value of generative AI, with EdTech organizations developing AI-based solutions to help students with their coursework and giving ideas for lessons to educators \cite{Kshetri_2023}. \citeauthor{Kshetri_2023} found that Quizlet launched an AI-Tutor Q-Chat, which combines ChatGPT with Quizlet's educational content library \cite{Kshetri_2023}. Furthermore, Khan Academy also started using AI to create a chat-bot, based on the GPT-4 model, with the goal in mind that students can use it to ask for assistance without the tool revealing the solution, but helping them solve the exercise \cite{Kshetri_2023}. 

The related work section shows that existing literature has explored traditional ITSs and the general capabilities of ChatGPT in various domains, and industry has already implemented sophisticated tools using generative AI. However, an exploratory understanding of its practical application as an AI-Tutor within APASs is still missing. The empirical studies have primarily focused on the model's ability to debug, generate code, and provide hints, and therefore, were able to state that large language models can be used for specific tasks, like tutoring. However, none have really implemented such a system and therefore, the real student experience, interaction patterns, and perceptions when using ChatGPT as an AI-Tutor in APASs have not been scientifically investigated. 
This study seeks to offer a scientific evaluation on the integration a GPT-based AI-Tutor in APASs, demonstrating that its tutoring capabilities, as proposed in literature, can be realized in practice.

\section{Methodological Approach}

This study is guided by a set of research questions (RQ1, RQ2 and RQ3) that have been defined in \hyperlink{Introduction}{Section 1}. 
To address these research questions, we implemented a three-stage methodological approach:

\begin{enumerate}
    \item \textbf{Integration of the AI-Tutor within an APAS:} Initial integration into the Artemis platform \cite{krusche2018artemis, linhuber2023constructive}.
    \item \textbf{Practical application by students:} Students solved a specific programming task on the platform.
    \item \textbf{Exploratory survey:} A survey targeting students of the "Introduction to Programming" course at the University of Innsbruck to collect their experiences with the AI-Tutor.
\end{enumerate}

The following sections explain the details of each of these stages.

\subsection{Integration of the AI-Tutor within an APAS}

We have implemented the AI-Tutor to collect data. This included developing a prototype that integrates the APAS Artemis \cite{krusche2018artemis, krusche2021interactive} with the GPT-3.5-Turbo model of OpenAI. 
We have chosen Artemis as the APAS for this study because of several reasons: 

\begin{enumerate}
	\item \textbf{Open Source}: Artemis is available as an open source project on GitHub, which makes this research reproducible.\footnote{\url{https://github.com/ls1intum/Artemis}}
	\item \textbf{Functional Scope}: Artemis provides all the basic features necessary for an APAS, including automatic exercise evaluation via test driven feedback, which improves the external validity of the findings \cite{sauerwein2023lecturers}.
	\item \textbf{Online Editor}: Artemis allows students to solve exercises online via a built-in code editor. This made the implementation of the AI-Tutor and data collection easier.
    \item \textbf{Large User Base}: Artemis is used by more than ten different universities, like the TU Munich and University of Innsbruck, and is therefore used by thousands of students every semester. Therefore, improving the Artemis APAS is directly beneficial for a large user base. 
\end{enumerate}

Before the integration of the AI-Tutor the workflow to use Artemis for programming exercises was the following \cite{krusche2018artemis}:

\begin{enumerate}
	\item \textbf{Instructors prepare an exercise}: Mainly involves the creation of an exercise description, the creation of a template file, the creation of a sample solution and the creation of unit tests to test the code submitted by the students.
	\item \textbf{Students solve an exercise}: Students write code to solve the problem statement using the integrated online editor offered by the platform. When students submit a solution attempt, the code of the submission is stored in a version control system. For this study, it was GitLab. 
	\item \textbf{System returns feedback}: For each submission, a build pipeline is triggered that executes the test cases written by the instructors and returns the test results with individual messages as feedback to the students.
\end{enumerate}

\begin{figure}
	\centering
	\includegraphics[width=\linewidth]{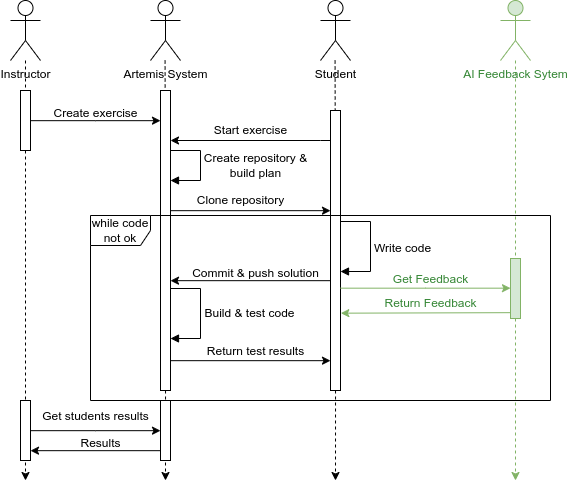}
	\caption{Sequence diagram of the usage workflow of the Artemis system extended by the AI-Tutor functionality.}
	\label{fig:system}
	\Description{Sequence diagram of the usage workflow of the Artemis system extended by the AI-Tutor functionality.}
\end{figure}

The integration of the AI-Tutor extended this workflow by an additional possibility to request feedback from the AI-Tutor. This extended workflow has been depicted in Figure \ref{fig:system}.

\begin{figure}
	\centering
	\includegraphics[width=\linewidth]{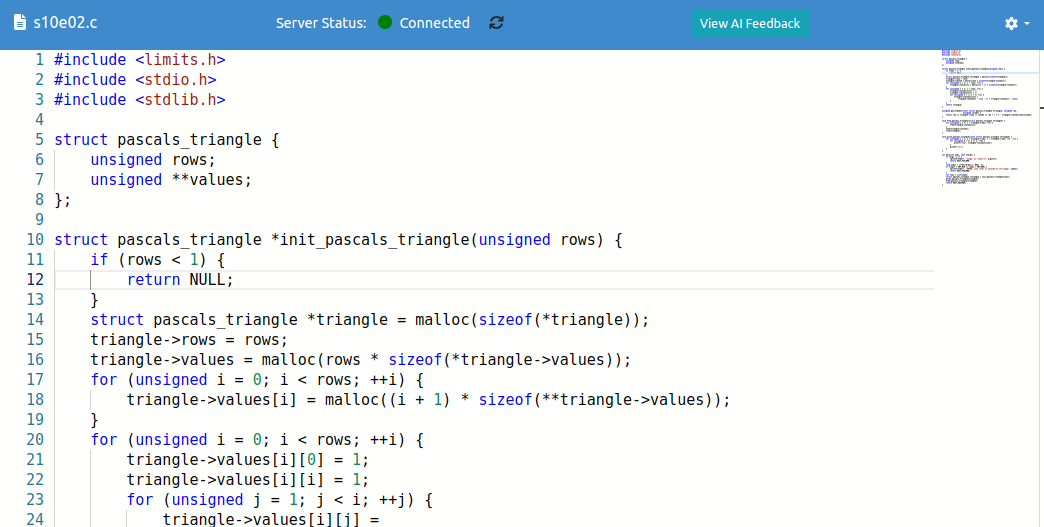}
	\caption{Artemis code editor with the button to "View AI Feedback".}
	\label{fig:screenshot001}
	\Description{Artemis code editor with the button to "View AI Feedback".}
\end{figure}

In Figure \ref{fig:screenshot001}, the Artemis code editor is shown with the new possibility to request AI feedback by clicking the "View AI Feedback" button displayed on the top right. 

Once this request is send to the server, the current solution of the student, the exercise description and sample solution are retrieved from the APAS' database and an API-call to the OpenAI servers containing the following information is sent:

\begin{enumerate}
	\item \textbf{Model}: The requested LLM, which is in our case GPT-3.5-Turbo.
	\item \textbf{Message}: The prompt which the LLM should take into consideration. This prompt can be seen in Listing \ref{lst:gptprompt}.
	\item \textbf{Temperature}: We set the temperature at 0.7 to balance predictability and creativity in the LLM's responses. This level ensures relevant feedback with sufficient variability for exploring diverse solutions.\footnote{\url{https://platform.openai.com/docs/guides/text-generation/how-should-i-set-the-temperature-parameter}}
\end{enumerate}

\begin{figure}
	\centering
	\includegraphics[width=\linewidth]{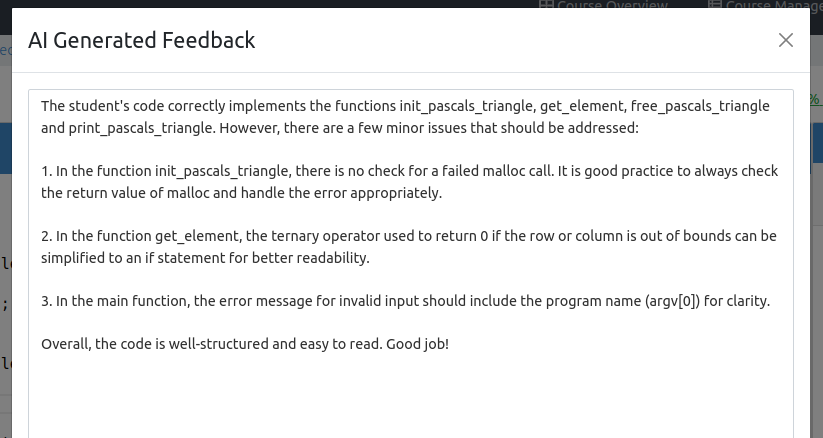}
	\caption{AI Generated Feedback displayed in a pop up.}
	\label{fig:screenshot002}
	\Description{AI Generated Feedback displayed in a pop up.}
\end{figure}

Once having sent this API-call to the OpenAI servers, we extracted the response of the LLM and displayed it in a pop-up window without any further modifications. This can be seen in Figure \ref{fig:screenshot002}. 
All the code files needed to adapt Artemis to display this possibility can be found in Figshare\footnote{\url{https://figshare.com/s/636a9c5ff8f2c8315f26}}.  

The language model receives the following prompt depicted in Listing \ref{lst:gptprompt}: 

\begin{lstlisting}[language=Java, caption={GPT-3.5-Turbo Prompt}, frame=single, basicstyle=\small, breaklines=true, escapeinside={\%*}{*)}, label=lst:gptprompt]
Act as a programming tutor and give informal
feedback in %*\textbf{language}*) to the student. 
The exercise description is the following: %*\textbf{description}*) 
The students code looks like that at the moment: %*\textbf{current}*)  
Do not provide a code solution. 
The optimal solution should look like that: %*\textbf{solution}*)
Important: Do not provide code.
\end{lstlisting}

The prompt starts with the main instruction to tell the LLM to act as a tutor.
\texttt{language} is the current selected language in Artemis. This value can either be English or German. 

\texttt{description} is the the task to be solved by the student. For this study the students had to implement Pascal's triangle and the task was to implement the functions to generate, display and release the memory for a portion of Pascal's triangle \cite{hinz1992pascal}.
In addition to the task description, the students were given a starting template with method stubs to start the exercise with. 

We have chosen the Pascal's triangle exercise to test the AI-Tutor because:

\begin{enumerate}
    \item \textbf{Foundational programming constructs}: The implementation of Pascal's triangle touches upon many foundational programming concepts such as loops, conditionals, arrays, and in some languages, dynamic memory management. If an AI can give valuable feedback on this exercise, it indicates its capability to understand and instruct on tasks involving the foundational concepts mentioned before.
    
    \item \textbf{Algorithmic thinking}: The process for creating Pascal's triangle involves iterative and recursive thinking. This showcases the AI's capability to handle a diverse range of algorithmic challenges as many other algorithm problems involve similar patterns of thought.
    
    \item \textbf{Concept overlap}: Many problems in computer science and mathematics share concepts with Pascal's triangle, e.g., the binomial expansion and combinatorics. A successful tutoring here indicates the AI's potential to generalize its capabilities to related problems.
    
    \item \textbf{Versatility in problem complexity}: Pascal's triangle can be approached in multiple ways. If the AI-tutor can manage the range of solutions for this problem, it suggests its robustness in tutoring exercises with different levels of complexity.
    
    \item \textbf{Debugging and problem solving}: Common mistakes are possible in implementing Pascal's triangle. An AI-tutor's ability to diagnose and correct these signifies its potential to generalize this capability to other programming challenges.

\end{enumerate}

The last two parts of the prompt, \texttt{current} and \texttt{solution}, represent the current solution of the student and the optimal solution defined by the exercise creator, respectively.

To integrate the AI-Tutor we have chosen to not enable direct interactions with the model because of the following reasons: 

\begin{enumerate}
	\item \textbf{Easier usability}: The assumption was that predetermining the prompt would simplify the user experience. By eliminating the chat-bot-style interaction, we sidestepped the necessity for students to formulate a question, thus streamlining their interactions.
	\item \textbf{Controlled environment}: A predefined interaction model provides a more controlled setting, thereby simplifying measures taken to avoid students from receiving the solution for the exercise via prompt engineering \cite{liu2023jailbreaking}.
	\item \textbf{Quality assurance}: With a static model, we were able to ensure that the AI-Tutor offers consistent pedagogically sound feedback, which is in line with the course's learning objectives.
	\item \textbf{Data privacy}: Direct interactions could inadvertently lead students to input sensitive or personal information. A static model minimizes this risk, adhering better to data privacy standards.
	\item \textbf{Resource efficiency}: Direct, dynamic interactions with the system may consume more resources, because the chat history should be given as context to enable meaningful conversations. Therefore, leading to higher costs as more tokens are used.
	\item \textbf{Reduction of over-reliance}: By limiting direct interactions, students were encouraged to think critically and not over-rely on the AI-Tutor for every minor query or challenge.
\end{enumerate}

When the student presses the button "View AI Feedback" we store the following information in the database: 
\begin{enumerate}
	\item \textbf{Code}: The current solution of the student.
	\item \textbf{Feedback}: The feedback returned by the LLM.
	\item \textbf{User}: A user identifier to identify each request.
	\item \textbf{File}: The file on which the student is working on.
	\item \textbf{Timestamp}: The current time.
\end{enumerate}

\subsection{Exploratory Survey}

We selected students from the "Introduction to Programming" tutorial as subjects for the experiment. This tutorial is part of the Bachelor in Computer Science curriculum at the University of Innsbruck and teaches first year students the basics in the programming language C. For this experiment a total of 23 students actively participated. While this may seem like a modest sample size, it's important to note that the qualitative nature of this analysis allowed for a more in-depth understanding of individual experiences, making the size not only manageable but also advantageous and given their recent interactions with traditional, human tutors, these students were especially appropriate subjects for assessing an AI-Tutor.

Prior to the data collection and survey implementation, we introduced the students to the possibility of receiving feedback from the newly implemented AI-Tutor via Artemis.

In the subsequent tutorial, the students were tasked with solving the, before described, Pascal's Triangle task \cite{hinz1992pascal}. They had one week to solve the task.
While they were solving the exercise they were free to choose whether they used the new AI-Feedback functionality or not. However, when they pressed the "View AI-Feedback"-button we stored the "AI Feedback data" (Code at feedback time, feedback returned by the LLM, User, File and Timestamp) in the database and when they submitted their current code to the Artemis system their current solution, test results and timestamp were saved in the version control system connected to Artemis. 

Finally, for the next course, we allotted approximately 15 minutes for the students to complete a questionnaire. 
In the survey we asked the following questions based on the Technology Acceptance Model (TAM)\cite{Davis1985}. TAM is a theoretical model that includes two primary factors that determine an individual's intention to use a technology: (1) Perceived Ease of Use (PEOU) and (2) Perceived Usefulness (PU). The model has been widely adopted in various fields to understand and predict the acceptance of newly implemented features.

\begin{enumerate}
    \item \textbf{I find the AI-Tutor easy to use}: This is directly related to the PEOU dimension of TAM. It seeks to collect the respondents' perceptions about the ease of interface and interaction with the AI-Tutor.
    \item \textbf{Using the AI-Tutor for my tasks enables me to accomplish the tasks more quickly}: This question is mainly about the efficiency offered by the AI-Tutor, which can be seen as a subset of PU as it implies the benefit of time-saving.
    \item \textbf{Using the AI-Tutor improves my performance}: This touches on the PU dimension by gauging whether the users feel they perform better in their tasks due to the AI-Tutor.
    \item \textbf{Using the AI-Tutor for my tasks increases my productivity}: Again, this is a question about PU. By increasing productivity, the AI-Tutor is seen as adding value to the user’s.
    \item \textbf{Using the AI-Tutor makes it easier to do my tasks}: This question is about both PEOU and PU. On one hand, it assesses ease of task accomplishment (PEOU), and on the other, it speaks to the utility value of the AI-Tutor (PU).
    \item \textbf{I find the AI-Tutor useful}: This is a direct reflection of the PU dimension, asking the respondent to evaluate the overall usefulness of the AI-Tutor.
\end{enumerate}

To obtain additional feedback, we asked the following open questions:

\begin{enumerate}
    \item What challenges did you encounter when utilizing the AI-Tutor?
    \item Do you have any further suggestions on how the AI-Tutor could be improved?
\end{enumerate}

Both questions aim to uncover specific difficulties or obstacles that users have faced, helping to identify specific areas for improvement in the design or functionality of the AI-Tutor.
Lastly, we asked questions about their demographics, including their highest degree, their current semester and their programming experience.  

\subsection{Data Analysis}

The data analysis involved first the combination of two datasets: The data saved when AI-Feedback was requested and the data saved when the students submitted their solutions. The "AI Feedback" dataset provided insights into the code at feedback time, feedback from the AI model, user details and timestamps. The student submissions included the code at submission time, the test results and the respective timestamps.

For accuracy, students who did not solve the exercise and did not engage with the AI-Tutor were excluded from the qualitative analysis allowing for an assessment of a total of 12 participants. This analysis was designed to identify patterns, and insights from the student responses, ensuring a comprehensive understanding of their experiences with the AI-Tutor.
This analysis included:

\begin{enumerate}
    \item \textbf{Temporal Coding:} Students' submissions and feedback request times were identified and marked in different colors to identify interaction patterns of students.
    \item \textbf{Thematic Coding:} Students' responses from the open-ended questions were initially read and re-read to identify common themes and patterns.
    \item \textbf{Theme Development:} The patterns were grouped under broader thematic categories, and a narrative was constructed around each theme. This involved interpreting the data within the context of this study's research questions.
\end{enumerate}

Key insights derived from this qualitative analysis were essential in understanding the intricacies of the student interactions and experiences. 

\section{Results}

In the following, we present the results of the conducted analysis. We divided this section into three subsections, each addressing a specific research question.

\subsection{Student Interaction}

Overall, the following interaction patterns emerged from the analysis of the data.
Four students neither made submissions to Artemis nor sought feedback from the AI-Tutor. One student made a single submission to Artemis without asking for any feedback from the AI-Tutor. In contrast, a different student sought feedback from the AI-Tutor once, yet did not submit anything to Artemis. Two students made several submissions to Artemis without seeking feedback from the AI-Tutor. Different patterns were observed, with one student displaying each of the following behaviors: making a single submission to Artemis and seeking feedback from the AI-Tutor once, making multiple submissions to Artemis and asking for feedback from the AI-Tutor once, and making a single submission to Artemis while seeking feedback from the AI-Tutor multiple times. It is particularly noteworthy that 12 students worked intensively with both systems, uploaded numerous submissions to Artemis and frequently asked for feedback from the AI tutor.

\begin{figure*}
    \centering
    \includegraphics[width=1\linewidth]{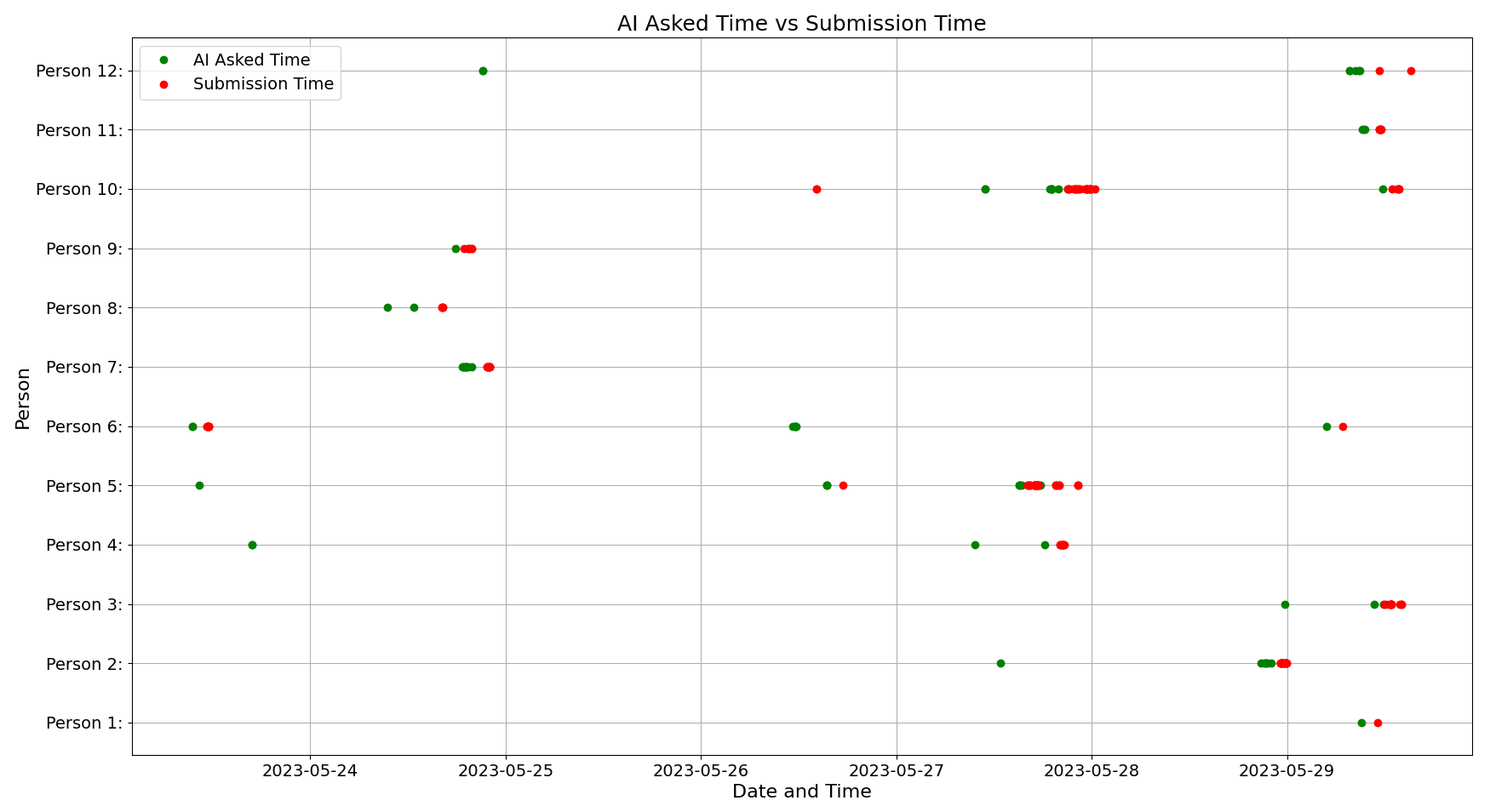}
    \caption{This figure shows the times when each student asked the AI-Tutor or submitted a code solution to the APAS.}
    \label{fig:interaction}
    \Description{Times when each of highly active students asked the AI-Tutor or submitted a code solution to the APAS.}
\end{figure*}

Considering its significance, we primarily focus on the behavior of the 12 students who exhibited high interaction rates with both Artemis and the AI-Tutor.
Figure \ref{fig:interaction} illustrates the timestamps when the students asked the AI-Tutor or submitted their solution to the APAS. On the Y-Axis, each line corresponds to a student and the X-Axis can be interpreted as a timeline starting with 2023-05-23 and ending with 2023-05-29. The red points in this figure indicate the time at which a student submitted their code to APAS. indicate the exact time at which a student requested feedback from the AI-Tutor. This figure shows that there are mainly two different ways in which students interact with the AI tutor. Based on this timeline, we were able to derive two user personas.

\subsubsection{Continuous Feedback - Iterative Ivy}

Iterative Ivy represents students who utilize the AI-Tutor intensively before their initial submission to the APAS. These students often begin without a complete solution and turn to the AI-Tutor for guidance on understanding and solving the exercise. The AI-Tutor, in its capacity, guides through specific instructions encompassing aspects like function implementation, memory management, and value calculation. Over multiple feedback cycles, students refine their solutions. When the AI-Tutor's feedback shifts towards minor optimizations, students tend to transition to submitting their work to the APAS, aiming for a perfect score.
 
\subsubsection{Alternating Feedback - Hybrid Harry}

Hybrid Harry exemplifies students who alternate between the AI-Tutor feedback and APAS submissions throughout their coding process. Typically, they begin their tasks by seeking initial insights from the AI-Tutor even before submitting a solution. Some send repeated requests for feedback on the same code segment, indicating potential uncertainties or the need for more explicit guidance. These students tend to submit their work to the APAS after establishing a foundation of their code. Notably, the AI-Tutor recognized incomplete or non-functional implementations, which students corrected after being told so by the AI-Tutor.

\vspace{0.5em}
\begin{recommendation}{}{}
We identified two user personas: (1) Continuous Feedback - Iterative Ivy, who relied mainly on AI feedback before final submissions to APAS, and (2) Alternating Feedback - Hybrid Harry, who alternately used the AI-Tutor and APAS submissions throughout the process.
\end{recommendation}

\subsection{Student Experience}

\begin{figure*}
    \centering
    \includegraphics[width=1\linewidth]{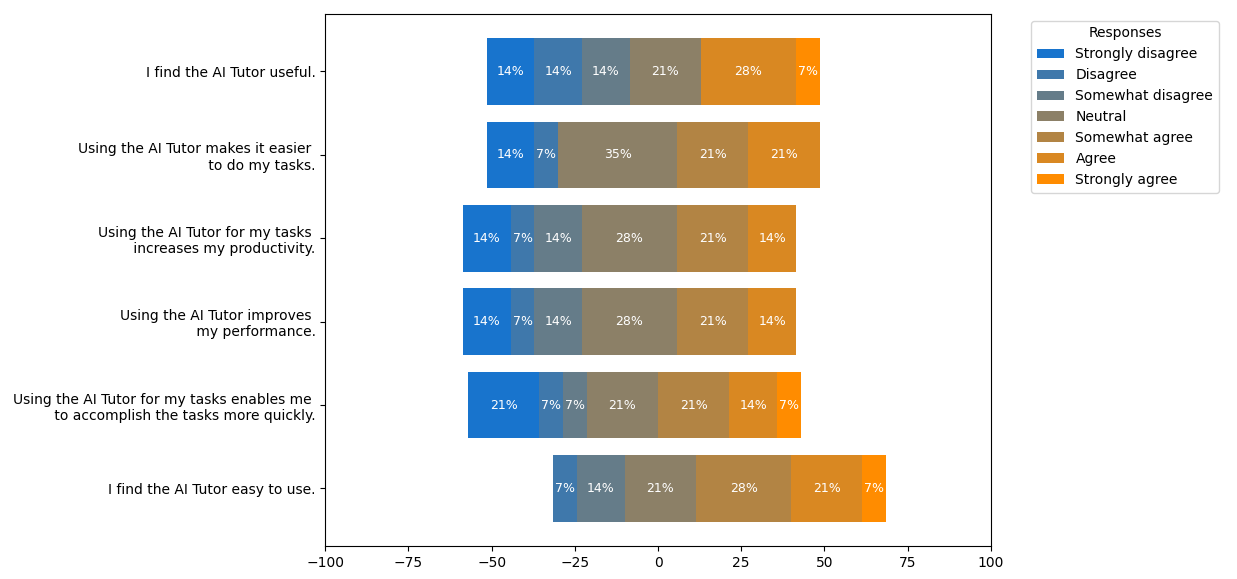}
    \caption{Students' satisfaction with the AI Tutor.}
    \label{fig:sentiment}
    \Description{This figure provides a visual representation of user sentiments towards the AI-Tutor, illustrating the relative occurrence of varying levels of agreement or disagreement with several statements, positioned based on average sentiment.}
\end{figure*}

Figure \ref{fig:sentiment} represents the distribution of user responses based on the questionnaire defined in Section 3. 
The responses are presented as horizontal stacked bars. Each bar represents a different statement, and the segments of the bar represent the proportion of responses for each level of agreement. The position of the bars along the X-axis reflects the average sentiment of the responses, ranging from negative on the left to positive on the right.
The zero point serves as a reference for interpreting Figure \ref{fig:sentiment}. If the majority of a bar lies to the left of this point, it generally indicates a more negative sentiment. Conversely, if it is situated to the right of the zero point, the sentiment is predominantly positive.  
Like this, the figure allows us to easily visualize how users perceive the AI-Tutor and its benefits.  
Examining this figure we found that reactions were mixed, ranging from positive to equally negative. However, the polarized responses appear to neutralize one another, resulting in a largely neutral overall response. 

Mapping the Likert scale from -3 to +3, using 0 as a neutral midpoint. The average sentiment for each answer is the following:

\begin{itemize}
    \item I find the AI Tutor easy to use: Somewhat Agree (1.29).
    \item Using the AI-Tutor for my tasks enables me to accomplish the tasks more quickly: Neutral (-0.29).
    \item Using the AI-Tutor improves my performance: Neutral (-0.43).
    \item Using the AI-Tutor for my tasks increases my productivity: Neutral (-0.43).
    \item Using the AI-Tutor makes it easier to do my tasks: Neutral (0.14).
    \item I find the AI-Tutor useful: Neutral (0.14).
\end{itemize}

Only the statement "I find the AI Tutor easy to use" received an other than neutral average response, indicating a mild agreement.

Regarding the open questions about challenges encountered while using Artemis and its AI-Tutor, student feedback consistently touched on several main themes:

\begin{enumerate}
    \item \textbf{Desire for Greater Specificity:} The AI-Tutor's responses were perceived as too generic. Students preferred more context-specific feedback pointing directly to improvement areas in the code.
    
    \item \textbf{Request for Increased Interactivity and Interface Concerns:} Students expressed the wish for enhanced interactive capabilities with the AI-Tutor, such as the ability to ask follow-up questions after initial feedback. Additionally, the interface was criticized because there was no possibility to see old feedback because once the feedback window was closed one could only request new feedback. 
    
    \item \textbf{Demand for Concrete Examples:} To supplement the written feedback, students believed that concrete code examples would help them interpreting the AI's suggestions.
    
    \item \textbf{Apprehension about Learning Inhibition:} Some students feared that using the AI-Tutor might lead to over-reliance which would slow down their learning progress.
\end{enumerate}

In terms of general feedback, students mentioned the system's potential, but also that it is notably perceived as an early-stage prototype. Furthermore, they compared its current utility to rudimentary software aids, expressing hope for more refined, context-aware feedback in future iterations.

\vspace{0.5em}
\begin{recommendation}{}{}
Some students found the system useful others stated the opposite resulting in an overall neutral result regarding the TAM. However, answers to the open-questions revealed that students, which gave mostly negative responses found the feedback to be too generic and lacking concrete examples.
\end{recommendation}

\subsection{Lessons Learned}

The practical integration of a large language model into an APAS offered valuable insights into the system's strengths and weaknesses. Through this experience, several key lessons and actionable insights can be derived.

A key lesson learned is that the AI-Tutor exhibits the capacity for real-time, personalized feedback provision. We have found that the system tends to return a more high level explanation of the task, if the students had not yet written a lot of code.
When the student has already written much code that is mostly correct, then the system tends to start giving recommendations on how to improve the code quality. For example, to add comments explaining the code or to change ternary operators to if-statements for better readability. 
An other surprising insight is that the AI-Tutor was able to give feedback on logical and semantic issues. We found that if students had defined wrong boundary conditions to terminate a loop, then the AI-Tutor recognized this and proposed to the student to change this condition.
This immediate feedback helped students to quickly correct their errors and thereby mitigating the acquisition of poor coding practices. Additionally, the system's inherent ability to serve feedback to a large, diverse student population in real-time underscores its applicability in large-scale educational contexts, particularly in Massive Open Online Courses (MOOCs).

However, we also learned that AI-Tutoring does not come without its challenges. The AI-Tutor, while efficient, occasionally delivered only general feedback, which means that there’s room to refine its responses for more detailed, code-specific guidance. 
Analyzing the feedback provided by the AI-Tutor we found that from 75 feedback requests 55(66.6\%) were useful and 20(26.6\%) were categorized as not useful. Among the 20 not useful responses we found that three revealed the solution of the exercise to be solved, four answers were hallucinations and 13 were too general to be helpful in the students situation. 
Mostly, if the answers were too general, we found that the AI-Tutor explained the exercise to the student even-though his or her solutions was already very sophisticated. The hallucinations were mainly about the AI-Tutor stating that a function looks well implemented, although there was no student implementation there yet or that a function should be implemented that was already implemented by the student.  

Additionally, as the existing system did not allow any interaction, we learned that enhancing its capability to address follow-up queries would improve the learning experience. Regarding the operational dependency, we found that downtimes in the API could jeopardize the tutor’s functionality. 

We also found that it is important to address students' over-reliance concerns, encouraging them to use the AI-Tutor without the fear that their learning progress might be hindered. Additionally, the feedback quality might get compromised due to context limits of models like GPT-3.5-Turbo. Exploring ways to manage this limitation effectively will be beneficial.
Despite careful prompt crafting, there were instances where the AI-Tutor revealed solutions. Ensuring that the model maintains adherence to the guidelines is a pivotal lesson.

\vspace{0.5em}
\begin{recommendation}{}{}
Implementing an AI-Tutor in an APAS showed that AI can support human tutors, allowing them to focus on deeper personal interactions. However, improvements are needed, as the AI's feedback was effective only 66.6\% of the time, being too generic, revealing solutions, or incorrect. Additionally, some students worry that using the AI-Tutor may slow their learning progress.
\end{recommendation}

\section{Discussion}

In this study we found mainly two usage strategies adopted by students when interacting with the AI-Tutor. It is important to understand these user personas, as they provide insights that can inform the design of AI-powered educational systems to be able to fit different learning styles and strategies.
First of all, we defined the user persona called Iterative Ivy. Users assigned to this persona first used the AI-Tutor intensively and only when the solution was already very advanced started to submit their solutions to the system. This approach seems to favor a traditional, linear programming strategy: (1) Comprehending the problem, (2) Writing a complete solution, and then (3) Validating the solution. By seeking continuous feedback from the AI-Tutor, these students ensured they were progressing on the right track before submitting their final solution. This finding suggests that AI-Tutors are beneficial to students who prefer to seek guidance and validation throughout their learning process, rather than just at the end. However, a potential concern here could be an over-reliance on the AI-Tutor. The continuous feedback-seeking behavior may stem from uncertainty or lack of confidence, which needs to be addressed in further pedagogical planning.
Secondly, we defined the user persona Hybrid Harry. Whose strategy contrasts sharply with Iterative Ivy's approach. Users assigned to this persona alternated between seeking AI-Tutor feedback and submitting solutions to the APAS. These students opted for an iterative learning approach, which represents a more agile programming practice. This suggests that AI-Tutors and APAS can facilitate active, self-regulated learning. However, the risk here lies in the potential for students to rely too heavily on the test feedback to guide their work, which could inadvertently lead to a trial-and-error approach to solve an exercise, rather than understanding the core principles.

Furthermore, given the responses to the TAM questions we found that students have mixed feelings regarding the usefulness of the AI-Tutor. While some students do not appreciate the help of the AI-Tutor others do appreciate it. It is important to identify the exact reasons why this is the case, but a first analysis indicates that the main reasons for the negative responses are user interface and prompt related. 
Students who responded negatively to the TAM questions also complained about the inability to ask follow-up questions and that the system did not return code examples or that the feedback is too general. The problem regarding the inability to ask follow up questions can be solved by changing to a chat-bot based system. The second problem regarding the missing code examples can be addressed by changing the prompt to allow code examples as responses. As a result, it is reasonable to conclude that GPT-3.5-Turbo can be successfully used as a language model behind an AI-Tutor. 

Addressing fears that AI-Tools may inhibit learning success is also crucial. Reiterating the tool's purpose to supplement rather than supplant traditional learning methods may decrease such concerns. 

Additionally, the use of AI-Tutors in programming education, as seen in this research, presents a unique set of learned lessons.
Regarding ITS we have found that LLMs offer a distinct adaptability advantage, because in conventional ITS, altering feedback mechanisms often demands intricate changes in the system's codebase, which can be time-consuming and resource-intensive. However, with LLMs, modifications are primarily done through prompt engineering. Given their vast training data, refining or adjusting the prompts can quickly adapt the feedback the model provides, without needing to change its internal mechanics. This allows educators to swiftly adapt to changing educational needs or methodologies.

Among the advantages of AI-Tutors over traditional human tutors, the promptness of feedback provided by AI-Tutors stands out as a game-changer. The capacity to instantly identify and correct errors can be vital for students learning programming, because quick feedback can minimize the propagation of misunderstandings and bad coding practices. Moreover, the scalability and cost-effectiveness of an AI-Tutor makes it a compelling choice, especially in resource-limited settings or with a large student base.

However, these benefits come with their share of challenges. An aspect of the AI-Tutor that needs attention is its occasional inclination to "hallucinate", producing responses that might not be entirely accurate or relevant. In this study, this primarily manifested when a student had already implemented a correct solution, resulting in the LLM sometimes advising the student to refine a program that was already functioning correctly. A potential mitigation strategy could involve integrating the AI feedback with unit test results. This would inform the student when their solution meets all criteria, signaling that subsequent AI feedback may not be entirely accurate.

The dependence on API availability and the inherent token limit of models like GPT-3.5-Turbo add an additional layer of complexity. Any change or downtime in the API can hinder the AI-Tutor's operations, and the context limitation imposed by the token limit can affect the quality of feedback, especially for more complex or lengthy code submissions.  

A more psychological perspective brings forth concerns about the impact of AI-Tutors on students' learning progress and the lack of a personal touch. An over-reliance on the AI-Tutor might impede students from developing their problem-solving skills, as they might rely too heavily on instant feedback rather than trying to debug and solve problems themselves. Additionally, the impersonal nature of an AI-Tutor might make the learning experience less engaging and less adaptive to individual student needs, which could affect motivation and learning outcomes.

These challenges should not overshadow the immense potential that an AI-Tutor holds. By addressing the issues mentioned, we can create more sophisticated and effective AI-Tutors that can significantly enhance the educational experience and outcomes for students learning programming. The journey towards optimizing AI-Tutors for programming education is still in progress, but the destination seems promising.

\section{Limitations}

In this paper we mainly considered the following four categories of validity, also used by \cite{wohlin2012experimentation}: (1) Construct validity, (2) Reliability, (3) Internal validity and (4) External validity.

\subsection{Construct Validity}

The research questions were defined using the PICOC system \cite{petticrew2008systematic}. The PICOC framework provides a systematic way to formulate research questions by emphasizing five elements: Population, Intervention, Comparison, Outcome, and Context. This structured approach ensures that research questions are both comprehensive and relevant. For this study:

\begin{enumerate}
\item \textbf{RQ1} primarily addresses the Population, Intervention, Outcome, and Context by examining the nature of student interactions within the specific context of an APAS assisted by an AI-Tutor.
\item \textbf{RQ2} focuses on the Population, Intervention, and Outcome by probing the students' experiences with AI-driven feedback when guided by the AI-Tutor.
\item \textbf{RQ3} encompasses all the PICOC elements, especially Context, by analyzing the broader lessons learned from deploying an AI-Tutor within the APAS environment.
\end{enumerate}

The research questions were further refined through discussions with several experts in the field to ensure alignment with the topic of interest. Leveraging the PICOC system as a foundation, coupled with the structured data collection approach and exploratory survey, facilitated a thorough answering of RQ1--3.

\subsection{Reliability}

We conducted a systematic data collection and analysis approach, as detailed in Section 3. Therefore, the process is both transparent and reproducible. 
However, it's crucial to note that the use of GPT-3.5-Turbo introduced a variable element. Given the nature of LLMs, not every prompt produces identical responses on different occasions. As a result, while the core structure and methodology can be reproduced, there may be slight variations in the responses generated by the model across different replications of the experiment.

\subsection{External Validity}

One potential limitation in this domain arises from the fact that we integrated the AI-Tutor only into Artemis. However, a systematic comparison of various APASs confirmed that Artemis' basic functionalities are echoed in many other APASs, deeming it a representative system \cite{sauerwein2023lecturers}. Additionally, the integration of the AI-Tutor can be done platform independent, because the approach stays the same, as it should be possible on all APAS to integrate a pop-up window that displays the results of the REST API calls.

Another potential threat to external validity is the use of a GPT model as the foundation for the AI-Tutor. While the used model is a state-of-the-art LLM and exhibits advanced conversational abilities, it might not perfectly mimic every possible LLM's behavior. Nonetheless, given that the model is based on the same foundational architectures as most other prevalent LLMs, and shares many of their characteristics and capabilities, we argue that the findings related to GPT-3.5-Turbo can largely be extrapolated to other similar models. It serves as a representative example, providing insights that are likely applicable across various LLMs.

Furthermore, the total number of participants can influence the external validity. In this study 23 students from the course "Introduction to Programming" participated. This sample size is too small to conduct a statistically significant quantitative analysis. As a result, we decided to focus on a qualitative analysis and report the experiences of implementing and operating an AI-Tutor. This allows for a deeper exploration of the students' experiences and behaviors when using the system. Last but not least, given the students' recent engagements with traditional human tutors, they were especially well-suited to evaluate the AI-Tutor.

\subsection{Internal Validity}

This study study largely leaned on qualitative analysis, which can sometimes introduce subjective bias. Nevertheless, the methodological rigor employed aimed to minimize such biases.
The detailed procedures involved in the qualitative analysis have been outlined in the research methodology section. By closely following these methodological steps, we have aimed to ensure that the findings are both credible and trustworthy. 

\section{Conclusion}

In this study of integrating the model behind ChatGPT as an AI-Tutor into the Artemis APAS, we uncovered both the immense potential and challenges of such an application. While the AI-Tutor offered advantages like timely feedback and scalability, its limitations were apparent. These included occasionally generic feedback, lack of interactive dialog, operational vulnerabilities related to API availability, potential over-reliance by students, the absence of a human touch, and technical constraints like context limits.

The vast potential of AI-Tutors in programming education is undeniable, but careful implementation and ongoing refinement are essential. This exploration underscores the need for more research in this domain, balancing technological progress with the irreplaceable human aspect of education.

Future work should focus on enhancing AI-Tutor's feedback specificity, interactivity improvements, user interface refinements, and addressing the token limit and prompt engineering challenges. The potential exploration of more powerful models like GPT-4 may further improve the feedback quality. This study's findings serve as a foundation for continued research in this innovative intersection of AI and education.

\section{Data Availability}

The data supporting the findings of this study are openly available in Figshare \footnote{\url{https://figshare.com/s/636a9c5ff8f2c8315f26}}. The dataset comprises the following:

\begin{enumerate}
    \item \textbf{Data Analysis ANONYM.xlsx}:
    \begin{enumerate}
        \item \textit{Sheet 1}: Contains extracted data from the database, such as Code, Feedback, User, Time, as well as various descriptive statistics, detailing, for instance, the frequency with which each user consulted the AI-Tutor, submissions to the APAS, and the final score.
        \item \textit{Sheet 2}: Houses the responses from the qualitative survey.
        \item \textit{Sheet 3}: Features an analysis that groups submissions and the state of the code when querying the AI-Tutor. It assesses the quality of the feedback and observes code alterations post-feedback.
    \end{enumerate}
    \item \textbf{Student submissions to the version control system}: Comprises multiple anonymized folders, each storing the code a student uploaded to the system. This code is augmented at the end with annotations detailing if the student had previously consulted the AI-Tutor and, if so, the associated timestamps.
    \item \textbf{Artemis-Files}: Contains all essential files to be integrated into your public Artemis project to activate the AI-Tutor functionality. These files are designed to work with the open-source Artemis project.\footnote{\url{https://github.com/ls1intum/Artemis}} A comprehensive repository has not been released due to challenges associated with its anonymization.
\end{enumerate}

It is essential to note that all personal identifiers have been removed to maintain confidentiality and adhere to data protection principles.

\section{Acknowledgments}

The CodeAbility Austria project has been funded by
the Austrian Federal Ministry of Education, Science and
Research (BMBWF).

\bibliographystyle{ACM-Reference-Format}
\bibliography{bibliography}


\begin{thebibliography}{34}


\ifx \showCODEN    \undefined \def \showCODEN     #1{\unskip}     \fi
\ifx \showDOI      \undefined \def \showDOI       #1{#1}\fi
\ifx \showISBNx    \undefined \def \showISBNx     #1{\unskip}     \fi
\ifx \showISBNxiii \undefined \def \showISBNxiii  #1{\unskip}     \fi
\ifx \showISSN     \undefined \def \showISSN      #1{\unskip}     \fi
\ifx \showLCCN     \undefined \def \showLCCN      #1{\unskip}     \fi
\ifx \shownote     \undefined \def \shownote      #1{#1}          \fi
\ifx \showarticletitle \undefined \def \showarticletitle #1{#1}   \fi
\ifx \showURL      \undefined \def \showURL       {\relax}        \fi
\providecommand\bibfield[2]{#2}
\providecommand\bibinfo[2]{#2}
\providecommand\natexlab[1]{#1}
\providecommand\showeprint[2][]{arXiv:#2}

\bibitem[Anderson and Skwarecki(1986)]%
        {Anderson_1986}
\bibfield{author}{\bibinfo{person}{J.~R Anderson} {and} \bibinfo{person}{E.
  Skwarecki}.} \bibinfo{year}{1986}\natexlab{}.
\newblock \showarticletitle{The automated tutoring of introductory computer
  programming}.
\newblock \bibinfo{journal}{\emph{Commun. ACM}} \bibinfo{volume}{29},
  \bibinfo{number}{9} (\bibinfo{date}{sep} \bibinfo{year}{1986}),
  \bibinfo{pages}{842--849}.
\newblock
\urldef\tempurl%
\url{https://doi.org/10.1145/6592.6593}
\showDOI{\tempurl}


\bibitem[Brade{\v{s}}ko and Mladeni{\'c}(2012)]%
        {bradevsko2012survey}
\bibfield{author}{\bibinfo{person}{Luka Brade{\v{s}}ko} {and}
  \bibinfo{person}{Dunja Mladeni{\'c}}.} \bibinfo{year}{2012}\natexlab{}.
\newblock \showarticletitle{A survey of chatbot systems through a loebner prize
  competition}. In \bibinfo{booktitle}{\emph{Proceedings of Slovenian language
  technologies society eighth conference of language technologies}},
  Vol.~\bibinfo{volume}{2}. sn, \bibinfo{pages}{34--37}.
\newblock


\bibitem[Brusilovsky(1992)]%
        {brusilovsky1992intelligent}
\bibfield{author}{\bibinfo{person}{PL Brusilovsky}.}
  \bibinfo{year}{1992}\natexlab{}.
\newblock \showarticletitle{Intelligent tutor, environment and manual for
  introductory programming}.
\newblock \bibinfo{journal}{\emph{Educational \& Training Technology
  International}} \bibinfo{volume}{29}, \bibinfo{number}{1}
  (\bibinfo{year}{1992}), \bibinfo{pages}{26--34}.
\newblock


\bibitem[Crow et~al\mbox{.}(2018)]%
        {Crow_2018}
\bibfield{author}{\bibinfo{person}{Tyne Crow}, \bibinfo{person}{Andrew
  Luxton-Reilly}, {and} \bibinfo{person}{Burkhard Wuensche}.}
  \bibinfo{year}{2018}\natexlab{}.
\newblock \showarticletitle{Intelligent tutoring systems for programming
  education}. In \bibinfo{booktitle}{\emph{Proceedings of the 20th Australasian
  Computing Education Conference}}. \bibinfo{publisher}{{ACM}}.
\newblock
\urldef\tempurl%
\url{https://doi.org/10.1145/3160489.3160492}
\showDOI{\tempurl}


\bibitem[Daun and Brings(2023)]%
        {daun2023chatgpt}
\bibfield{author}{\bibinfo{person}{Marian Daun} {and} \bibinfo{person}{Jennifer
  Brings}.} \bibinfo{year}{2023}\natexlab{}.
\newblock \showarticletitle{How ChatGPT Will Change Software Engineering
  Education}. In \bibinfo{booktitle}{\emph{Proceedings of the Conference on
  Innovation and Technology in Computer Science Education V. 1}}.
  \bibinfo{pages}{110--116}.
\newblock


\bibitem[Davis(1985)]%
        {Davis1985}
\bibfield{author}{\bibinfo{person}{Fred~D. Davis}.}
  \bibinfo{year}{1985}\natexlab{}.
\newblock \emph{\bibinfo{title}{A Technology Acceptance Model for Empirically
  Testing New End-User Information Systems: Theory and Results}}.
\newblock \bibinfo{thesistype}{Ph.\,D. Dissertation}.
  \bibinfo{school}{Massachusetts Institute of Technology, Sloan School of
  Management}.
\newblock


\bibitem[Ghimire et~al\mbox{.}(2020)]%
        {ghimire2020accelerating}
\bibfield{author}{\bibinfo{person}{Awishkar Ghimire},
  \bibinfo{person}{Surendrabikram Thapa}, \bibinfo{person}{Avinash~Kumar Jha},
  \bibinfo{person}{Surabhi Adhikari}, {and} \bibinfo{person}{Ankit Kumar}.}
  \bibinfo{year}{2020}\natexlab{}.
\newblock \showarticletitle{Accelerating business growth with big data and
  artificial intelligence}. In \bibinfo{booktitle}{\emph{2020 Fourth
  International Conference on I-SMAC (IoT in Social, Mobile, Analytics and
  Cloud)(I-SMAC)}}. IEEE, \bibinfo{pages}{441--448}.
\newblock


\bibitem[Hinz(1992)]%
        {hinz1992pascal}
\bibfield{author}{\bibinfo{person}{Andreas~M Hinz}.}
  \bibinfo{year}{1992}\natexlab{}.
\newblock \showarticletitle{Pascal's Triangle and the Tower of Hanoi}.
\newblock \bibinfo{journal}{\emph{The American mathematical monthly}}
  \bibinfo{volume}{99}, \bibinfo{number}{6} (\bibinfo{year}{1992}),
  \bibinfo{pages}{538--544}.
\newblock


\bibitem[Ho et~al\mbox{.}(2020)]%
        {ho2020enabling}
\bibfield{author}{\bibinfo{person}{Dean Ho}, \bibinfo{person}{Stephen~R Quake},
  \bibinfo{person}{Edward~RB McCabe}, \bibinfo{person}{Wee~Joo Chng},
  \bibinfo{person}{Edward~K Chow}, \bibinfo{person}{Xianting Ding},
  \bibinfo{person}{Bruce~D Gelb}, \bibinfo{person}{Geoffrey~S Ginsburg},
  \bibinfo{person}{Jason Hassenstab}, \bibinfo{person}{Chih-Ming Ho},
  {et~al\mbox{.}}} \bibinfo{year}{2020}\natexlab{}.
\newblock \showarticletitle{Enabling technologies for personalized and
  precision medicine}.
\newblock \bibinfo{journal}{\emph{Trends in biotechnology}}
  \bibinfo{volume}{38}, \bibinfo{number}{5} (\bibinfo{year}{2020}),
  \bibinfo{pages}{497--518}.
\newblock


\bibitem[Holland et~al\mbox{.}(2009)]%
        {holland2009j}
\bibfield{author}{\bibinfo{person}{Jay Holland}, \bibinfo{person}{Antonija
  Mitrovic}, {and} \bibinfo{person}{Brent Martin}.}
  \bibinfo{year}{2009}\natexlab{}.
\newblock \showarticletitle{J-LATTE: a Constraint-based Tutor for Java}.
\newblock  (\bibinfo{year}{2009}).
\newblock


\bibitem[Jalil et~al\mbox{.}(2023)]%
        {jalil2023chatgpt}
\bibfield{author}{\bibinfo{person}{Sajed Jalil}, \bibinfo{person}{Suzzana
  Rafi}, \bibinfo{person}{Thomas~D LaToza}, \bibinfo{person}{Kevin Moran},
  {and} \bibinfo{person}{Wing Lam}.} \bibinfo{year}{2023}\natexlab{}.
\newblock \showarticletitle{Chatgpt and software testing education: Promises \&
  perils}. In \bibinfo{booktitle}{\emph{2023 IEEE International Conference on
  Software Testing, Verification and Validation Workshops (ICSTW)}}. IEEE,
  \bibinfo{pages}{4130--4137}.
\newblock


\bibitem[Kasneci et~al\mbox{.}(2023)]%
        {Kasneci2023chatgpt}
\bibfield{author}{\bibinfo{person}{Enkelejda Kasneci}, \bibinfo{person}{Kathrin
  Sessler}, \bibinfo{person}{Stefan Küchemann}, \bibinfo{person}{Maria
  Bannert}, \bibinfo{person}{Daryna Dementieva}, \bibinfo{person}{Frank
  Fischer}, \bibinfo{person}{Urs Gasser}, \bibinfo{person}{Georg Groh},
  \bibinfo{person}{Stephan Günnemann}, \bibinfo{person}{Eyke Hüllermeier},
  \bibinfo{person}{Stephan Krusche}, \bibinfo{person}{Gitta Kutyniok},
  \bibinfo{person}{Tilman Michaeli}, \bibinfo{person}{Claudia Nerdel},
  \bibinfo{person}{Jürgen Pfeffer}, \bibinfo{person}{Oleksandra Poquet},
  \bibinfo{person}{Michael Sailer}, \bibinfo{person}{Albrecht Schmidt},
  \bibinfo{person}{Tina Seidel}, \bibinfo{person}{Matthias Stadler},
  \bibinfo{person}{Jochen Weller}, \bibinfo{person}{Jochen Kuhn}, {and}
  \bibinfo{person}{Gjergji Kasneci}.} \bibinfo{year}{2023}\natexlab{}.
\newblock \showarticletitle{{ChatGPT} for good? On opportunities and challenges
  of large language models for education}.
\newblock \bibinfo{journal}{\emph{Learning and Individual Differences}}
  \bibinfo{volume}{103} (\bibinfo{date}{apr} \bibinfo{year}{2023}),
  \bibinfo{pages}{102274}.
\newblock
\urldef\tempurl%
\url{https://doi.org/10.1016/j.lindif.2023.102274}
\showDOI{\tempurl}


\bibitem[Keuning et~al\mbox{.}(2016)]%
        {keuning2016towards}
\bibfield{author}{\bibinfo{person}{Hieke Keuning}, \bibinfo{person}{Johan
  Jeuring}, {and} \bibinfo{person}{Bastiaan Heeren}.}
  \bibinfo{year}{2016}\natexlab{}.
\newblock \showarticletitle{Towards a systematic review of automated feedback
  generation for programming exercises}. In
  \bibinfo{booktitle}{\emph{Proceedings of the 2016 ACM Conference on
  Innovation and Technology in Computer Science Education}}.
  \bibinfo{pages}{41--46}.
\newblock


\bibitem[Krusche(2021)]%
        {krusche2021interactive}
\bibfield{author}{\bibinfo{person}{Stephan Krusche}.}
  \bibinfo{year}{2021}\natexlab{}.
\newblock \emph{\bibinfo{title}{Interactive learning - A scalable and adaptive
  learning approach for large courses}}.
\newblock Habilitation. \bibinfo{school}{Technische Universität München}.
\newblock


\bibitem[Krusche and Seitz(2018)]%
        {krusche2018artemis}
\bibfield{author}{\bibinfo{person}{Stephan Krusche} {and}
  \bibinfo{person}{Andreas Seitz}.} \bibinfo{year}{2018}\natexlab{}.
\newblock \showarticletitle{Artemis: An automatic assessment management system
  for interactive learning}. In \bibinfo{booktitle}{\emph{Proceedings of the
  49th ACM technical symposium on computer science education}}.
  \bibinfo{pages}{284--289}.
\newblock


\bibitem[Kshetri(2023)]%
        {Kshetri_2023}
\bibfield{author}{\bibinfo{person}{Nir Kshetri}.}
  \bibinfo{year}{2023}\natexlab{}.
\newblock \showarticletitle{The Economics of Generative Artificial Intelligence
  in the Academic Industry}.
\newblock \bibinfo{journal}{\emph{Computer}} \bibinfo{volume}{56},
  \bibinfo{number}{8} (\bibinfo{date}{aug} \bibinfo{year}{2023}),
  \bibinfo{pages}{77--83}.
\newblock
\urldef\tempurl%
\url{https://doi.org/10.1109/mc.2023.3278089}
\showDOI{\tempurl}


\bibitem[Kuhail et~al\mbox{.}(2023)]%
        {kuhail2023interacting}
\bibfield{author}{\bibinfo{person}{Mohammad~Amin Kuhail},
  \bibinfo{person}{Nazik Alturki}, \bibinfo{person}{Salwa Alramlawi}, {and}
  \bibinfo{person}{Kholood Alhejori}.} \bibinfo{year}{2023}\natexlab{}.
\newblock \showarticletitle{Interacting with educational chatbots: A systematic
  review}.
\newblock \bibinfo{journal}{\emph{Education and Information Technologies}}
  \bibinfo{volume}{28}, \bibinfo{number}{1} (\bibinfo{year}{2023}),
  \bibinfo{pages}{973--1018}.
\newblock


\bibitem[Kulik and Fletcher(2016)]%
        {kulik2016effectiveness}
\bibfield{author}{\bibinfo{person}{James~A Kulik} {and} \bibinfo{person}{JD
  Fletcher}.} \bibinfo{year}{2016}\natexlab{}.
\newblock \showarticletitle{Effectiveness of intelligent tutoring systems: a
  meta-analytic review}.
\newblock \bibinfo{journal}{\emph{Review of educational research}}
  \bibinfo{volume}{86}, \bibinfo{number}{1} (\bibinfo{year}{2016}),
  \bibinfo{pages}{42--78}.
\newblock


\bibitem[Linhuber et~al\mbox{.}(2023)]%
        {linhuber2023constructive}
\bibfield{author}{\bibinfo{person}{Matthias Linhuber},
  \bibinfo{person}{Jan~Philip Bernius}, {and} \bibinfo{person}{Stephan
  Krusche}.} \bibinfo{year}{2023}\natexlab{}.
\newblock \showarticletitle{Constructive Alignment in Modern Computing
  Education: An Open-Source Computer-Based Examination System}. In
  \bibinfo{booktitle}{\emph{23nd Koli Calling International Conference on
  Computing Education Research}} (Koli, Finland) \emph{(\bibinfo{series}{{Koli}
  '23})}.
\newblock
\urldef\tempurl%
\url{https://doi.org/10.35542/osf.io/nmpf6}
\showDOI{\tempurl}


\bibitem[Liu et~al\mbox{.}(2023)]%
        {liu2023jailbreaking}
\bibfield{author}{\bibinfo{person}{Yi Liu}, \bibinfo{person}{Gelei Deng},
  \bibinfo{person}{Zhengzi Xu}, \bibinfo{person}{Yuekang Li},
  \bibinfo{person}{Yaowen Zheng}, \bibinfo{person}{Ying Zhang},
  \bibinfo{person}{Lida Zhao}, \bibinfo{person}{Tianwei Zhang}, {and}
  \bibinfo{person}{Yang Liu}.} \bibinfo{year}{2023}\natexlab{}.
\newblock \showarticletitle{Jailbreaking chatgpt via prompt engineering: An
  empirical study}.
\newblock \bibinfo{journal}{\emph{arXiv preprint arXiv:2305.13860}}
  (\bibinfo{year}{2023}).
\newblock


\bibitem[Lo(2023)]%
        {lo2023impact}
\bibfield{author}{\bibinfo{person}{Chung~Kwan Lo}.}
  \bibinfo{year}{2023}\natexlab{}.
\newblock \showarticletitle{What is the impact of ChatGPT on education? A rapid
  review of the literature}.
\newblock \bibinfo{journal}{\emph{Education Sciences}} \bibinfo{volume}{13},
  \bibinfo{number}{4} (\bibinfo{year}{2023}), \bibinfo{pages}{410}.
\newblock


\bibitem[Ouh et~al\mbox{.}(2023)]%
        {ouh2023chatgpt}
\bibfield{author}{\bibinfo{person}{Eng~Lieh Ouh}, \bibinfo{person}{Benjamin
  Kok~Siew Gan}, \bibinfo{person}{Kyong~Jin Shim}, {and}
  \bibinfo{person}{Swavek Wlodkowski}.} \bibinfo{year}{2023}\natexlab{}.
\newblock \showarticletitle{ChatGPT, Can You Generate Solutions for my Coding
  Exercises? An Evaluation on its Effectiveness in an undergraduate Java
  Programming Course}.
\newblock \bibinfo{journal}{\emph{arXiv preprint arXiv:2305.13680}}
  (\bibinfo{year}{2023}).
\newblock


\bibitem[Pardos and Bhandari(2023)]%
        {pardos2023learning}
\bibfield{author}{\bibinfo{person}{Zachary~A Pardos} {and}
  \bibinfo{person}{Shreya Bhandari}.} \bibinfo{year}{2023}\natexlab{}.
\newblock \showarticletitle{Learning gain differences between ChatGPT and human
  tutor generated algebra hints}.
\newblock \bibinfo{journal}{\emph{arXiv preprint arXiv:2302.06871}}
  (\bibinfo{year}{2023}).
\newblock


\bibitem[Petticrew and Roberts(2008)]%
        {petticrew2008systematic}
\bibfield{author}{\bibinfo{person}{Mark Petticrew} {and} \bibinfo{person}{Helen
  Roberts}.} \bibinfo{year}{2008}\natexlab{}.
\newblock \bibinfo{booktitle}{\emph{Systematic reviews in the social sciences:
  A practical guide}}.
\newblock \bibinfo{publisher}{John Wiley \& Sons}.
\newblock


\bibitem[Ray(2023)]%
        {ray2023chatgpt}
\bibfield{author}{\bibinfo{person}{Partha~Pratim Ray}.}
  \bibinfo{year}{2023}\natexlab{}.
\newblock \showarticletitle{ChatGPT: A comprehensive review on background,
  applications, key challenges, bias, ethics, limitations and future scope}.
\newblock \bibinfo{journal}{\emph{Internet of Things and Cyber-Physical
  Systems}} (\bibinfo{year}{2023}).
\newblock


\bibitem[Rudolph et~al\mbox{.}(2023)]%
        {rudolph2023chatgpt}
\bibfield{author}{\bibinfo{person}{J{\"u}rgen Rudolph}, \bibinfo{person}{Samson
  Tan}, {and} \bibinfo{person}{Shannon Tan}.} \bibinfo{year}{2023}\natexlab{}.
\newblock \showarticletitle{ChatGPT: Bullshit spewer or the end of traditional
  assessments in higher education?}
\newblock \bibinfo{journal}{\emph{Journal of Applied Learning and Teaching}}
  \bibinfo{volume}{6}, \bibinfo{number}{1} (\bibinfo{year}{2023}).
\newblock


\bibitem[Sauerwein et~al\mbox{.}(2023)]%
        {sauerwein2023lecturers}
\bibfield{author}{\bibinfo{person}{Clemens Sauerwein}, \bibinfo{person}{Simon
  Priller}, \bibinfo{person}{Martin Dobiasch}, \bibinfo{person}{Stefan Oppl},
  \bibinfo{person}{Michael Felderer}, {and} \bibinfo{person}{Ruth Breu}.}
  \bibinfo{year}{2023}\natexlab{}.
\newblock \showarticletitle{Lecturers’ and Students’ Experiences with an
  Automated Programming Assessment System}.
\newblock  (\bibinfo{year}{2023}).
\newblock


\bibitem[Sobania et~al\mbox{.}(2023)]%
        {sobania2023analysis}
\bibfield{author}{\bibinfo{person}{Dominik Sobania}, \bibinfo{person}{Martin
  Briesch}, \bibinfo{person}{Carol Hanna}, {and} \bibinfo{person}{Justyna
  Petke}.} \bibinfo{year}{2023}\natexlab{}.
\newblock \showarticletitle{An analysis of the automatic bug fixing performance
  of chatgpt}.
\newblock \bibinfo{journal}{\emph{arXiv preprint arXiv:2301.08653}}
  (\bibinfo{year}{2023}).
\newblock


\bibitem[Stutz et~al\mbox{.}(2023)]%
        {stutz2023ch}
\bibfield{author}{\bibinfo{person}{Petra Stutz}, \bibinfo{person}{Maximilian
  Elixhauser}, \bibinfo{person}{Judith Grubinger-Preiner},
  \bibinfo{person}{Vivienne Linner}, \bibinfo{person}{Eva
  Reibersdorfer-Adelsberger}, \bibinfo{person}{Christoph Traun},
  \bibinfo{person}{Gudrun Wallentin}, \bibinfo{person}{Katharina W{\"o}hs},
  {and} \bibinfo{person}{Thomas Zuberb{\"u}hler}.}
  \bibinfo{year}{2023}\natexlab{}.
\newblock \showarticletitle{Ch (e) atgpt? an Anecdotal Approach on the Impact
  of Chatgpt on Teaching and Learning Giscience}.
\newblock \bibinfo{journal}{\emph{Preprint) https://doi. org/10.35542/osf.
  io/j3m9b}} (\bibinfo{year}{2023}).
\newblock


\bibitem[Surameery and Shakor(2023)]%
        {surameery2023use}
\bibfield{author}{\bibinfo{person}{Nigar M~Shafiq Surameery} {and}
  \bibinfo{person}{Mohammed~Y Shakor}.} \bibinfo{year}{2023}\natexlab{}.
\newblock \showarticletitle{Use chat gpt to solve programming bugs}.
\newblock \bibinfo{journal}{\emph{International Journal of Information
  Technology \& Computer Engineering (IJITC) ISSN: 2455-5290}}
  \bibinfo{volume}{3}, \bibinfo{number}{01} (\bibinfo{year}{2023}),
  \bibinfo{pages}{17--22}.
\newblock


\bibitem[Tian et~al\mbox{.}(2023)]%
        {tian2023chatgpt}
\bibfield{author}{\bibinfo{person}{Haoye Tian}, \bibinfo{person}{Weiqi Lu},
  \bibinfo{person}{Tsz~On Li}, \bibinfo{person}{Xunzhu Tang},
  \bibinfo{person}{Shing-Chi Cheung}, \bibinfo{person}{Jacques Klein}, {and}
  \bibinfo{person}{Tegawend{\'e}~F Bissyand{\'e}}.}
  \bibinfo{year}{2023}\natexlab{}.
\newblock \showarticletitle{Is ChatGPT the Ultimate Programming Assistant--How
  far is it?}
\newblock \bibinfo{journal}{\emph{arXiv preprint arXiv:2304.11938}}
  (\bibinfo{year}{2023}).
\newblock


\bibitem[Winkler and S{\"o}llner(2018)]%
        {winkler2018unleashing}
\bibfield{author}{\bibinfo{person}{Rainer Winkler} {and}
  \bibinfo{person}{Matthias S{\"o}llner}.} \bibinfo{year}{2018}\natexlab{}.
\newblock \showarticletitle{Unleashing the potential of chatbots in education:
  A state-of-the-art analysis}. In \bibinfo{booktitle}{\emph{Academy of
  Management Proceedings}}, Vol.~\bibinfo{volume}{2018}. Academy of Management
  Briarcliff Manor, NY 10510, \bibinfo{pages}{15903}.
\newblock


\bibitem[Wohlin et~al\mbox{.}(2012)]%
        {wohlin2012experimentation}
\bibfield{author}{\bibinfo{person}{Claes Wohlin}, \bibinfo{person}{Per
  Runeson}, \bibinfo{person}{Martin H{\"o}st}, \bibinfo{person}{Magnus~C
  Ohlsson}, \bibinfo{person}{Bj{\"o}rn Regnell}, {and} \bibinfo{person}{Anders
  Wessl{\'e}n}.} \bibinfo{year}{2012}\natexlab{}.
\newblock \bibinfo{booktitle}{\emph{Experimentation in software engineering}}.
\newblock \bibinfo{publisher}{Springer Science \& Business Media}.
\newblock


\bibitem[Zhang and Lu(2021)]%
        {zhang2021study}
\bibfield{author}{\bibinfo{person}{Caiming Zhang} {and} \bibinfo{person}{Yang
  Lu}.} \bibinfo{year}{2021}\natexlab{}.
\newblock \showarticletitle{Study on artificial intelligence: The state of the
  art and future prospects}.
\newblock \bibinfo{journal}{\emph{Journal of Industrial Information
  Integration}}  \bibinfo{volume}{23} (\bibinfo{year}{2021}),
  \bibinfo{pages}{100224}.
\newblock


\end{thebibliography}

\end{document}